\documentclass{IEEEtran}
	\usepackage[switch]{lineno}
	\usepackage{cite}
	\usepackage{xcolor,soul,framed}
	\usepackage[pdftex]{graphicx}
		\DeclareGraphicsExtensions{.png}

	\usepackage{amsmath,amssymb,amsfonts}
	\usepackage{textcomp,nicefrac}
	\usepackage[hidelinks]{hyperref}

\begin{document}

\title{Value-assigned pulse shape discrimination for neutron detectors}
\author{
    F.C.E. Teh,
	J.-W. Lee,
	K. Zhu,
	K.W. Brown,
	Z. Chajecki,
	W.G. Lynch,
	M.B. Tsang,
	A. Anthony,
	J. Barney,\\
	D. Dell'Aquila,
	J. Estee,
	B. Hong,
	G. Jhang,
	O.B. Khanal,
	Y.J. Kim,
	H.S. Lee,
	J.W. Lee,
	J. Manfredi,\\
	S.H. Nam,
	C.Y. Niu,
	J.H. Park,
	S. Sweany,
	C.Y. Tsang,
	R. Wang,
	H. Wu
	\thanks{
		Manuscript received on February 18, 2021. This work is supported by the U.S. National Science Foundation under Grant No. PHY-1565546 and the U.S. Department of Energy (Office of Science) under Grant No. DE-NA0003908. This work is also supported by the National Research Foundation of Korea (NRF) under grant Nos. 2016K1A3A7A09005578, 2018R1A5A1025563 and 2013M7A1A1075764.
	}
	\thanks{
		F. C. E. Teh, K. Zhu, W. G. Lynch, M. B. Tsang, A. Anthony, J. Barney, J. Estee, S. Sweany and C. Y. Tsang are with the National Superconducting Cyclotron Laboratory and the Department of Physics and Astronomy, Michigan State University, East Lansing, MI 48824, USA. Corresponding authors are F. C. E. Teh (teh@nscl.msu.edu) and M. B. Tsang (tsang@nscl.msu.edu).
	}
	\thanks{
		J. -W. Lee, B. Hong, J. W. Lee, S. H. Nam and J. H. Park are with the Department of Physics, Korea University, Seoul 02841, Republic of Korea.
	}
	\thanks{
		K. W. Brown is with the National Superconducting Cyclotron Laboratory and the Department of Chemistry, Michigan State University, East Lansing, MI 48824, USA.
	}
	\thanks{
		Z. Chajecki and O. B. Khanal are with the Department of Physics, Western Michigan University, Kalamazoo, MI 49008, USA.
	}
	\thanks{
		D. Dell'Aquila was with the National Superconducting Cyclotron Laboratory, East Lansing, MI 48824, USA. He is now with Universit\`{a} degli Studi di Sassari, 07100 Sassari SS, Italy.
	}
	\thanks{
		G. Jhang, C. Y. Niu and R. Wang are with the National Superconducting Cyclotron Laboratory, East Lansing, MI 48824, USA.
	}
	\thanks{
		Y. J. Kim and H. S. Lee are with the Rare Isotope Science Project, Institute of Basic Science, Daejeon 34047, Republic of Korea.
	}
	\thanks{
		J. Manfredi was with the National Superconducting Cyclotron Laboratory and the Department of Physics and Astronomy, Michigan State University, East Lansing, MI 48824, USA. He is now with the Department of Nuclear Engineering, University of California, Berkeley, CA 94720, USA.
	}
	\thanks{
		H. Wu is with the School of Physics, Peking University, Beijing 100871, China.
	}
}

\maketitle

\markboth{Preprint to IEEE Transactions on Nuclear Science}{Preprint to IEEE Transactions on Nuclear Science}

\begin{abstract}
	Using the waveforms from a digital electronic system, an offline analysis technique on pulse shape discrimination (PSD) has been developed to improve the neutron-gamma separation in a bar-shaped NE-213 scintillator that couples to a photomultiplier tube (PMT) at each end. The new improved method, called the ``valued-assigned PSD'' (VPSD), assigns a normalized fitting residual to every waveform as the PSD value. This procedure then facilitates the incorporation of longitudinal position dependence of the scintillator, which further enhances the PSD capability of the detector system. In this paper, we use radiation emitted from an AmBe neutron source to demonstrate that the resulting neutron-gamma identification has been much improved when compared to the traditional technique that uses the geometric mean of light outputs from both PMTs. The new method has also been modified and applied to a recent experiment at the National Superconducting Cyclotron Laboratory (NSCL) that uses an analog electronic system.
\end{abstract}


\section{Introduction}
	Detector arrays composed of multiple long bar-shaped or cylindrical scintillators dedicated for fast neutron detection are increasingly common in the experimental studies of nuclear physics \cite{zecher1997large,kogler2013light,sakai1992construction,bravar2009calibration,heideman2019conceptual,stuhl2017pandora,blaich1992large,celano1997performance,baumann2005construction,perdikakis2009lenda,langer2011simulations,peters2016performance,shim2019performance,douma2019investigation,li2019prototype}. Emissions of neutrons are typically accompanied by other types of radiation including gamma rays and charged particles, most of which have larger interaction cross sections with the scintillation material than the neutrons. To discriminate charged particles, one can place a charged-particle veto wall in front of the neutron detector \cite{douma2019investigation} or apply an external magnetic field to deflect the charged particles \cite{otsu2016samurai,bird2005system} away from the neutron detectors. To distinguish neutrons and gamma rays with pulse shape discrimination (PSD), organic scintillators are commonly used, e.g. NE-213/BC-501/EJ-301 in \cite{zecher1997large,sakai1992construction,bravar2009calibration,kogler2013light}, EJ-276 in \cite{heideman2019conceptual} and EJ-299 in \cite{stuhl2017pandora}. PSD utilizes the difference in the ``slow'' component of scintillation signal which depends on the particle type to distinguish neutrons and gamma rays \cite{brooks1959scintillation,kuchnir1968time}. PSD can only be omitted in the experimental studies when the yields of gamma rays and cosmic rays relative to the neutron yield are insignificant and can be treated as subtractable background \cite{baumann2002improving}, allowing the use of plastic scintillators like BC-408/EJ-200 that offers no PSD capability \cite{blaich1992large,celano1997performance,baumann2005construction,perdikakis2009lenda,langer2011simulations,peters2016performance,shim2019performance,douma2019investigation}.

	In recent years, there have been many efforts to develop new scintillation materials that have better PSD characteristics \cite{grodzicka2020fast,zaitseva2018recent,yang2017li,bertrand2015pulse,pozzi2013pulse,glodo2012pulse}. Construction of a large area (e.g. $2~\mathrm{m}\times2~\mathrm{m}$) neutron detector with the new materials is currently out of reach mainly due to cost. On the other hand, with the advancement of digital electronics, storage of the digitized waveforms facilitates the exploration of better and innovative analysis techniques, pushing the limits of PSD of both old and new detectors \cite{nakhostin2019general,safari2016discrete,sosa2016comparison,balmer2015comparative,liu2010digital,liu2009investigation,d2007digital}. In this paper, we focus on analysis of digital waveforms of the scintillation signals from large neutron walls. By analyzing light signals digitized with a flash analog-to-digital converter (FADC), we explore the gating conditions of the waveforms and devise a modification to the PSD analysis procedure to better distinguish neutrons and gamma rays. We develop the procedure using a bar-shaped scintillator with a photomuliplier tube (PMT) for light collection attached on each of the two ends. However, the method should be applicable to any PSD analysis involving two signals such as the traditional ``fast'' (early component of the scintillation pulse) and ``slow'' signals from analog electronics. The development of new PSD techniques is particularly timely as the existing PSD methods \cite{zecher1997large,kogler2013light,heideman2019conceptual,sakai1992construction,bravar2009calibration,li2019prototype} do not work well for long scintillation bars that have significant attenuation effect. We will discuss this in detail in Section~\ref{ssec:III-B} and Section~\ref{ssec:III-C}.

\section{Experimental setup}
	A $2~\mathrm{m}\times2~\mathrm{m}$ large area neutron wall (LANA) \cite{zecher1997large,zhu2020calibration} consists of $25$ independent detection units has been used as the neutron detection system for heavy-ion collision (HIC) experiments at the National Superconducting Cyclotron Laboratory (NSCL). Each detection unit is called a ``neutron wall (NW) bar'', which stacks horizontally on top of one another with an inter-bar gap of $0.3~\mathrm{cm}$. The NW bar is made up of a $200~\mathrm{cm}$ long and $0.3~\mathrm{cm}$ thick Pyrex glass container that is filled with NE-213 organic liquid scintillator. Its rectangular cross section has a vertical height of $7.62~\mathrm{cm}$ and a depth of $6.35~\mathrm{cm}$. At the two ends of each NW bar, a $7.5~\mathrm{cm}$-diameter Philips Photonics XP4312B/04 PMT is coupled to the Pyrex glass for photon detection.

	In a recent experiment, $7$ out of $25$ NW bars in a LANA wall, namely, bars enumerated as 01, 02, 03, 10, 11, 12, 13, have their signals split into two: (1) $90\%$ of the pulse is sent to the analog electronic system \cite{zhu2020calibration}; (2) the remaining $10\%$ is sent to the digital electronic system described by \cite{shim2019performance}. Ideally, the $10\%$ signal should be amplified 10 times to recover the original pulse. The NSCL Quad Fast Amps used in the experiment have fixed gain that varies between $8$ to $10$ times across different channels. Unless otherwise specified, the analysis results shown in this paper are drawn from NW bar 01 for illustration. All NW bars show similar results. Calibration of hit position on the NW bar is done using cosmic muons \cite{zhu2020calibration,zaihong2012performance}. To develop an improved pulse shape discrimination technique, we use AmBe as our primary neutron source throughout the paper, except in Section \ref{ssec:hic} where we analyze around half a million of neutrons and gamma rays emitted from the nuclear collisions of $^{48}\mathrm{Ca} + ^{64}\mathrm{Ni}$ with beam energy at $140~\mathrm{MeV/u}$ have been processed with the analog electronics. The AmBe source in use has an activity of $4.6~\mathrm{Ci}$. It was placed around $60~\mathrm{cm}$ in front of the $2~\mathrm{m}\times2~\mathrm{m}$ LANA wall and exposed for about two hours, resulting in a total of $\sim4.22\times10^5$ events being recorded in NW bar 01, $48\%$ of those later are identified as neutrons with the new PSD technique we propose in this paper.

\section{Pulse shape discrimination}

\subsection{Pre-processing of digitized waveforms}
	Each digitized pulse consists of $240$ samples taken at a time interval of $2~\mathrm{ns}$ ($500~\mathrm{MHz}$), each with a $12$-bit analog-to-digital converter (ADC). In our experiment, the detected pulses have a rise time of about $4~\mathrm{ns}$, defined as the time difference for the pulse rising from $10\%$ to $90\%$ of its peak. Hence the choice of interpolation is critical if we wish to reliably infer any pulse information between two consecutive samples in order to determine the signal start time and charge integration. Cubic spline interpolation is favored over linear interpolation as it allows us to (1) reasonably reproduce any missing peak of a pulse due to discrete sampling, while (2) avoiding the Runge's phenomenon that yields unrealistic oscillation in high-order polynomial interpolation \cite{epperson1987runge}.

	As will be demonstrated later in the section \ref{ssec:III-E}, PSD using cubic spline interpolation offers better n--$\gamma$ separation when compared to linear interpolation. Similar improvement in time resolution has been reported \cite{modamio2015digital}. In the remaining parts of this paper, all characteristics of a pulse are drawn from its cubic spline interpolation unless otherwise specified. To avoid numerical issues in the subsequent analyses, the integer ADC values are smeared by adding a random number between $-0.5$ and $+0.5$ to each ADC value.

\subsection{Traditional method using geometric mean of the charge signals}
	\label{ssec:III-B}
	\begin{figure}
	\centering
	\includegraphics[width=\columnwidth]{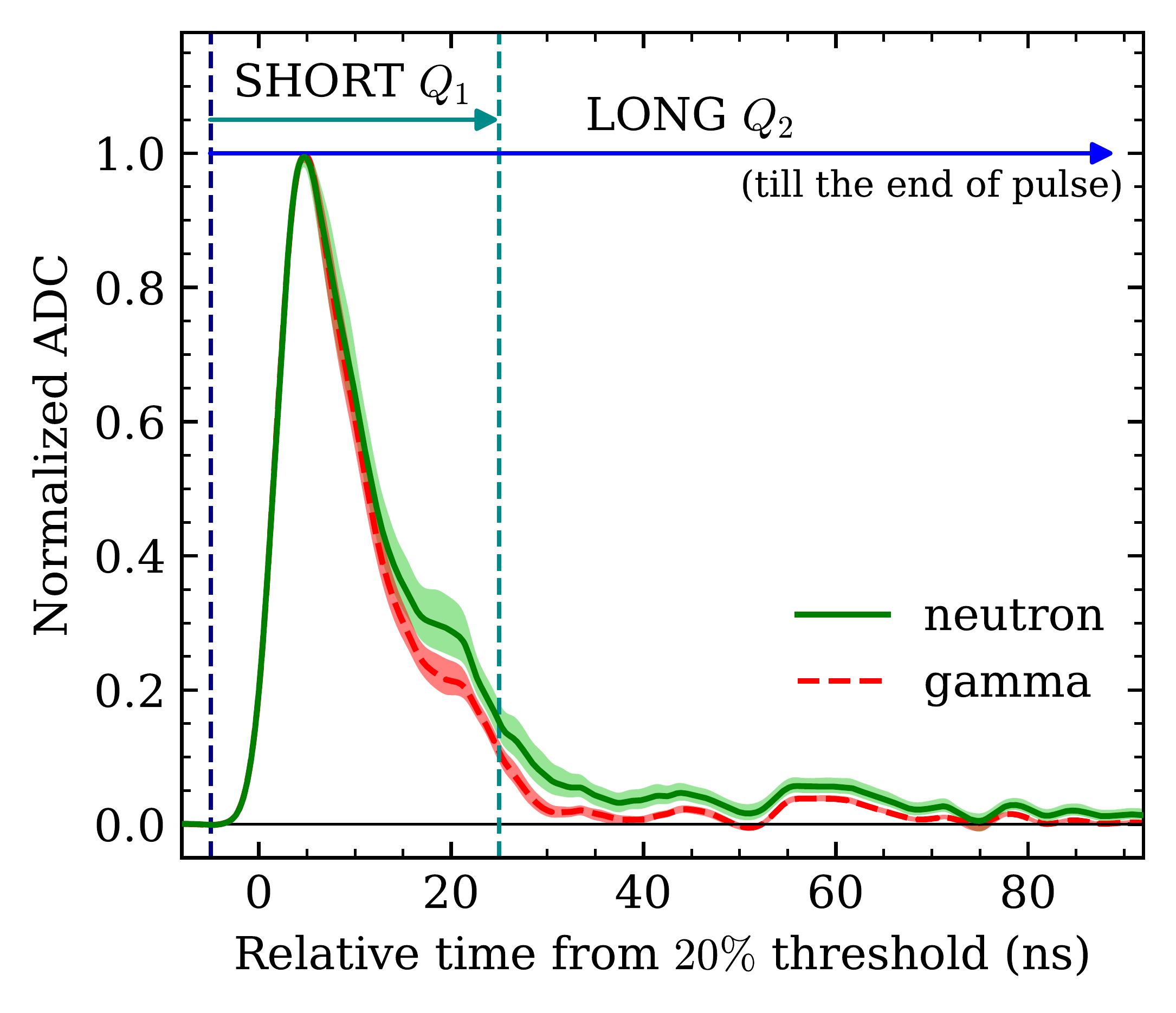}
	\caption{The cumulative waveforms detected by the left PMT of NW bar 01 with radiations from the AmBe source classified according to \figurename~\ref{fig:vpsd_2d}. Only hits on the closest $50~\mathrm{cm}$ (a quarter of the entire detector bar) to the left PMT have been included here. The shaded regions indicate the interquartile range of normalized ADC. Curves have all been interpolated with cubic spline.}
	\label{fig:median_pulse_shape}
	\end{figure}

	The n--$\gamma$ PSD is based on the pulse shape differences between neutrons and gamma rays. \figurename~\ref{fig:median_pulse_shape} shows the cumulative neutron pulses (solid) and gamma pulses (dashed), where the oscillating pattern at the tails is perhaps due to some cabling issues, particularly the splitting of signals into analog and digital electronic systems as well as amplification via the Fast Amps to recover signal strength. All of the difference in pulse shape only emerge after the peak. PSD is commonly achieved by analyzing the difference between two charge-integration gates over different time ranges \cite{zecher1997large,sakai1992construction,bravar2009calibration,kogler2013light,stuhl2017pandora,heideman2019conceptual,morris1976digital,ellis2017neutron,ivanova2016fast,janvcavr2015pulse,macmullin2013measurement}. There exist several nomenclatures in the literature for these two components. In this paper, they will be referred to as ``SHORT gate'' ($Q_1$) and ``LONG gate'' ($Q_2$) as illustrated in \figurename~\ref{fig:median_pulse_shape}. Mathematically, they are computed by integrating the digital pulse within the gate, i.e.
	\begin{linenomath*}
	\begin{equation}
		Q_{1, 2} = \int_{t^\mathrm{start}_{1, 2}}^{t^\mathrm{stop}_{1, 2}} S(t) \ dt \ ,
	\end{equation}
	\end{linenomath*}
	where $S(t)$ is the cubic spline interpolated waveform. \figurename~\ref{fig:median_pulse_shape} presents the optimal ranges of $Q_1$ and $Q_2$ on pulses observed in this work. Since the optimal constant time fraction differs among different NW bars, the pulses have all been aligned to the constant fraction timing of $20\%$ instead of the optimal one (e.g. $25\%$ for NW bar 01) to facilitate comparison among different NW bars; time resolution at constant fraction of $20\%$ differs less than $10~\mathrm{ps}$ from the optimal one.

	\begin{figure}
	\centering
	\includegraphics[width=\columnwidth]{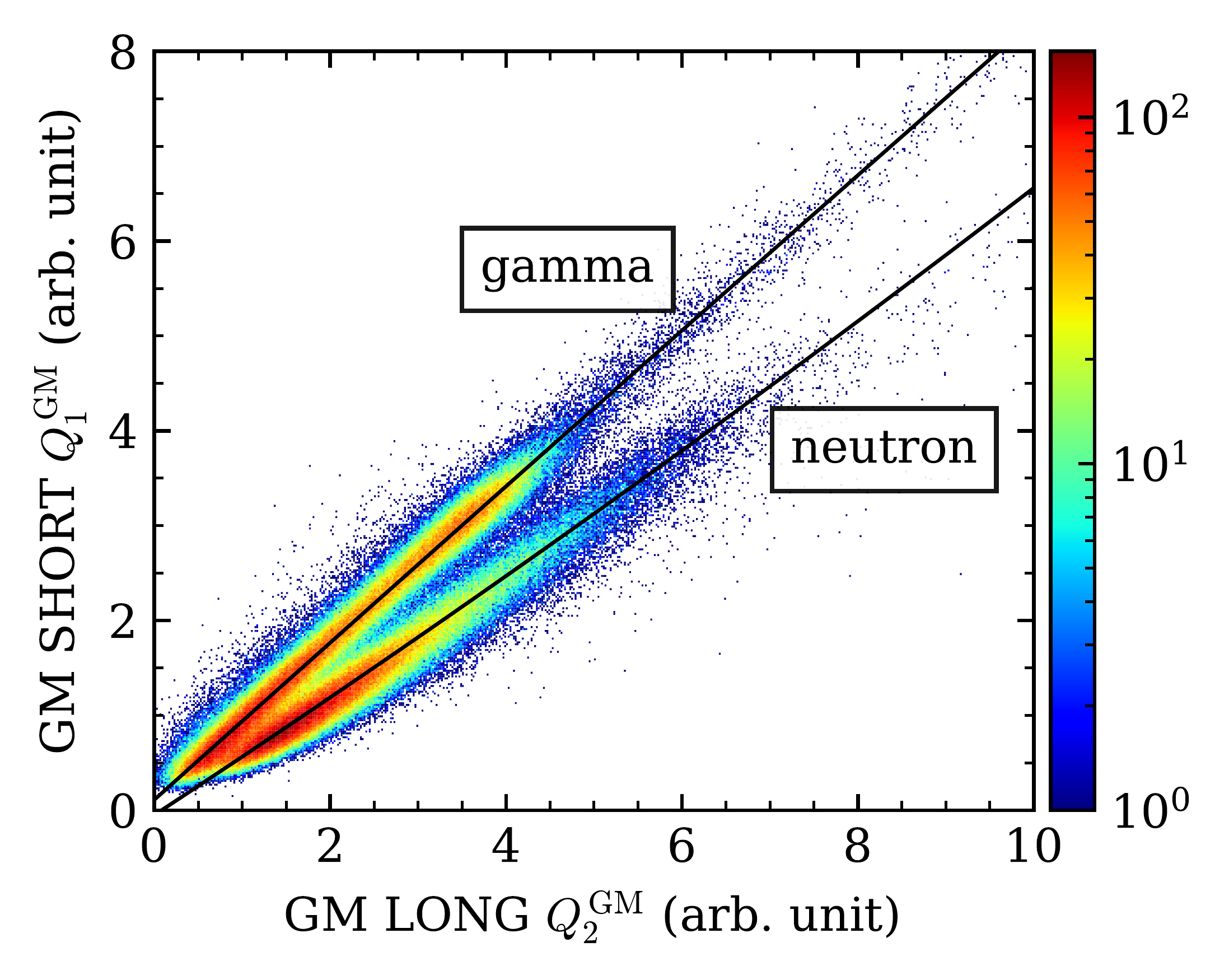}
	\caption{Traditional PSD obtained by plotting $Q^\mathrm{GM}_1$ vs. $Q^\mathrm{GM}_2$. The two ridges correspond to gamma rays and neutrons from the AmBe source.}
	\label{fig:geomean_2d}
	\end{figure}

	For a long scintillation detector, it is important to minimize the position dependence of signals, which is typically accomplished by combining signals from both PMTs of one NW bar. The traditional approach to utilize signals from two ends is to take the geometric mean (GM) of light outputs from two PMTs, $Q^\mathrm{GM}\equiv\sqrt{Q^\mathrm{L}Q^\mathrm{R}}$, where $Q^\mathrm{L}$ and $Q^\mathrm{R}$ are the light outputs from the left side and the right side, respectively \cite{zecher1997large,kogler2013light,heideman2019conceptual,sakai1992construction,bravar2009calibration,li2019prototype}. The GM signals may reduce position dependence that is mainly caused by the attenuation of light in the scintillation material. To see this, we denote $\lambda$ as the attenuation length and $\ell$ as the total length of a neutron wall bar. Then the light outputs detected by both PMTs on a NW bar would be given by
	\begin{linenomath*}
	\begin{equation}
		\begin{cases}
			Q^\mathrm{L}(x) = Q_0e^{-(\ell/2+x)/\lambda} \\
			Q^\mathrm{R}(x) = Q_0e^{-(\ell/2-x)/\lambda}
		\end{cases}
		\ ,
	\end{equation}
	\end{linenomath*}
	where $Q_0$ is the ideal light output assuming zero attenuation and $x$ is the distance from the center of NW bar to the position of interaction; our convention sets the right PMT at $x = \ell/2 = +100~\mathrm{cm}$. By taking the geometric mean of $Q^\mathrm{L}$ and $Q^\mathrm{R}$, i.e.
	\begin{linenomath*}
	\begin{equation}
		Q^\mathrm{GM} \equiv \sqrt{Q^\mathrm{L}Q^\mathrm{R}} = Q_0e^{-\ell/\lambda} \ ,
		\label{eq:GM}
	\end{equation}
	\end{linenomath*}
	we may recover the ideal light output $Q_0$ up to a position-independent factor of $e^{-\ell/\lambda}$. To extract $\lambda$, we notice
	\begin{linenomath*}
	\begin{equation}
		\ln\left(\frac{Q^\mathrm{L}(x)}{Q^\mathrm{R}(x)}\right) = -\frac{2}{\lambda}x \ .
	\end{equation}
	\end{linenomath*}
	Hence, $\lambda$ can be determined from the slope of $\ln(Q^\mathrm{L}/Q^\mathrm{R})$ as a function of $x$, and it is measured to be $90~\mathrm{cm}$ \cite{zhu2020calibration}.

	In this paper, this PSD technique that uses  the geometric means of signals is referred as the ``GM method'', and its performance is shown in \figurename~\ref{fig:geomean_2d} which serves as a benchmark for the development of an improved method.

\subsection{Value-assigned pulse shape discrimination (VPSD)}
	\label{ssec:III-C}

	\begin{figure*}
	\centering
	\includegraphics[width=\textwidth]{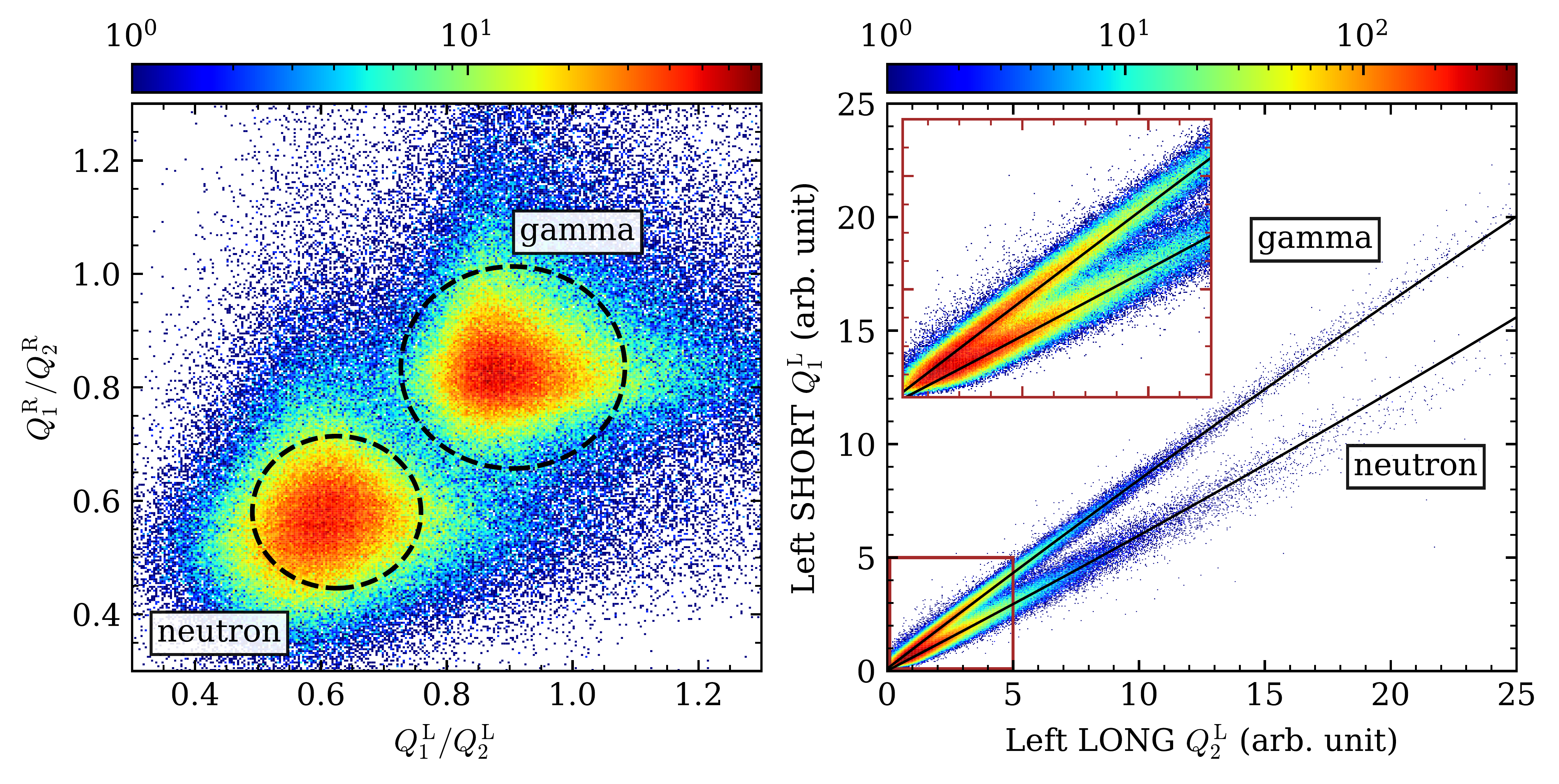}
	\caption{Left: PSD observed in a two-dimensional histogram of $Q_1/Q_2$ ratios measured at both ends. Right: A two-dimensional histogram of $(Q_2,Q_1)$ from the left end of NW bar 01. The two solid curves in the right panel are quadratic fits on the points gated by the dashed circles on the left panel. Both are radiations from the AmBe source.}
	\label{fig:ratio_2d}
	\end{figure*}

	While the GM method can minimize the position dependence, its construction averages out the individual performance of both PMTs. To recover contributions from individual PMTs, we plot a two-dimensional histogram of $Q_1/Q_2$ ratios measured by the PMTs at both ends of NW bar 01 in the left panel of \figurename~\ref{fig:ratio_2d}. Visually, separation of neutrons from gamma rays in the left panel of \figurename~\ref{fig:ratio_2d} is superior to that shown in \figurename~\ref{fig:geomean_2d}. Retaining information on individual PMT produces better PSD than to construct the geometric mean quantities. This simple plot of two-dimensional $Q_1/Q_2$ ratios obtained from PMTs at both ends is in contrast with the common practice, where the ratio $Q_1^\mathrm{GM} / Q_2^\mathrm{GM}$ is being used \cite{zecher1997large,kogler2013light,heideman2019conceptual,li2019prototype}. A similar technique that uses the $Q_1/Q_2$ ratios was applied in \cite{stuhl2017pandora} for a much shorter bar ($30~\mathrm{cm}$). The separation of n--$\gamma$ is comparable to our work. However, the earlier work did not correct for position dependence, which is essential for long scintillators.

	To further improve the PSD and to accommodate the position correction in the next subsection, we account for the slight nonlinearity by fitting the two ridges corresponding to neutrons and gamma rays in the right panel of \figurename~\ref{fig:ratio_2d}. We find that quadratic curves can reasonably model the $Q_1$--$Q_2$ relations of neutrons and gamma rays over the range of interest. Selective gates such as the dashed circles in the left panel can be applied to ensure better fitting of the ridges in the right panel.

	\begin{figure}
	\centering
	\includegraphics[width=\columnwidth]{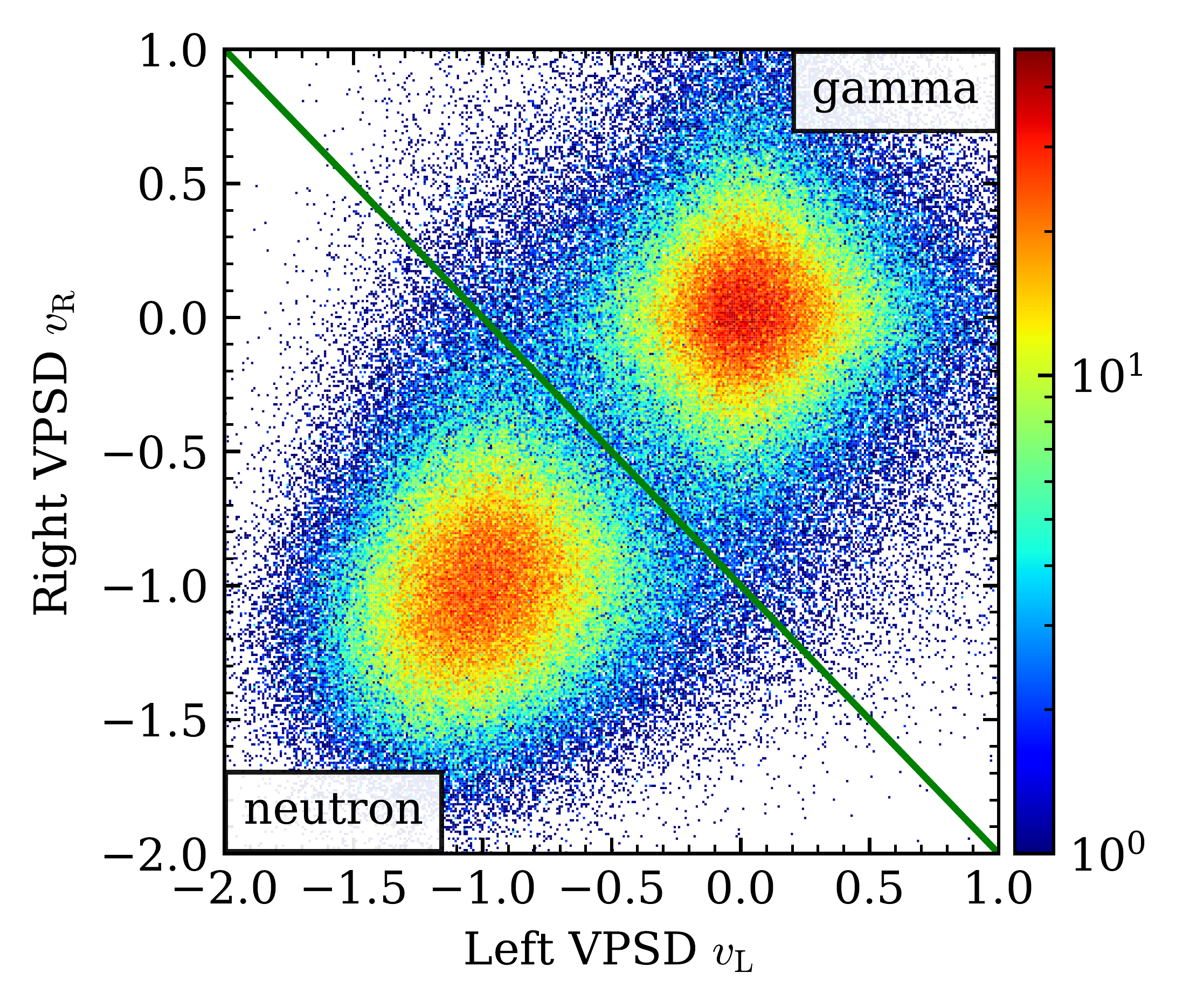}
	\caption{Two-dimensional histogram of VPSDs from the same set of data in \figurename~\ref{fig:ratio_2d}, AmBe source. As expected from the definition of $v_{\mathrm{L},\mathrm{R}}$, the clusters are centered near $(-1,-1)$ and $(0,0)$ for neutron and gamma, respectively. The solid diagonal line can be used for simple n--$\gamma$ classification.}
	\label{fig:vpsd_2d}
	\end{figure}

	\begin{table}
	\centering
	\caption{Fit parameters of $Q_1$--$Q_2$ relations for NW bar 01}
	\label{tab:fit}
	{\renewcommand{\arraystretch}{1.5}
	\begin{tabular}{| c | c c c |}
		\hline
		$\times 10^{-2}$      & $a_0$  & $a_1$  & $a_2$ \\
		\hline
		$q_n^\mathrm{L}$      & $3.90\pm0.10$ & $59.94\pm0.04$ & $-0.040\pm0.003$ \\
		$q_\gamma^\mathrm{L}$ & $12.35\pm0.08$ & $84.42\pm0.03$ & $-0.068\pm0.001$ \\
		$q_n^\mathrm{R}$      & $1.25\pm0.10$ & $56.82\pm0.04$ & $0.017\pm0.002$ \\
		$q_\gamma^\mathrm{R}$ & $6.58\pm0.08$ & $80.67\pm0.02$ & $-0.064\pm0.001$ \\
		\hline
	\multicolumn{4}{p{220pt}}{Fit parameters of $Q_1 \equiv q(Q_2) = a_0 + a_1 Q_2 + a_2 Q_2^2$. The right panel of \figurename~\ref{fig:ratio_2d} shows the fitted $Q_1$--$Q_2$ relations for the left PMT. Relations for the right PMT are not shown, but they are calculated in a similar fashion.}
	\end{tabular}
	}
	\end{table}

	The fit parameters for NW bar 01 are presented in Table~\ref{tab:fit}. Having found the quadratic functions for neutron ($q_n$) and gamma ($q_\gamma$), we may then assign a value for each pulse according to
	\begin{linenomath*}
	\begin{equation}
		v_i(Q_1, Q_2) \equiv \frac{Q^i_1-q_\gamma^i(Q^i_2)}{q_\gamma^i(Q^i_2) - q_n^i(Q^i_2)}
	\end{equation}
	\end{linenomath*}
	for each side, $i = \mathrm{L}, \mathrm{R}$. This quantity $v$ shall be called the ``valued-assigned pulse shape discrimination'' (VPSD). A two-dimensional histogram of VPSDs for a NW bar is shown in \figurename~\ref{fig:vpsd_2d}, in which the shapes of both clusters have become more globular when compared to the left panel of \figurename~\ref{fig:ratio_2d} because the quadratic fits have accounted for the nonlinearity in the $Q_1$--$Q_2$ relations. Compared to the GM method as shown in \figurename~\ref{fig:geomean_2d}, the PSD has improved drastically.

\subsection{Position dependence of VPSD}
\begin{figure}
	\centering
	\includegraphics[width=\columnwidth]{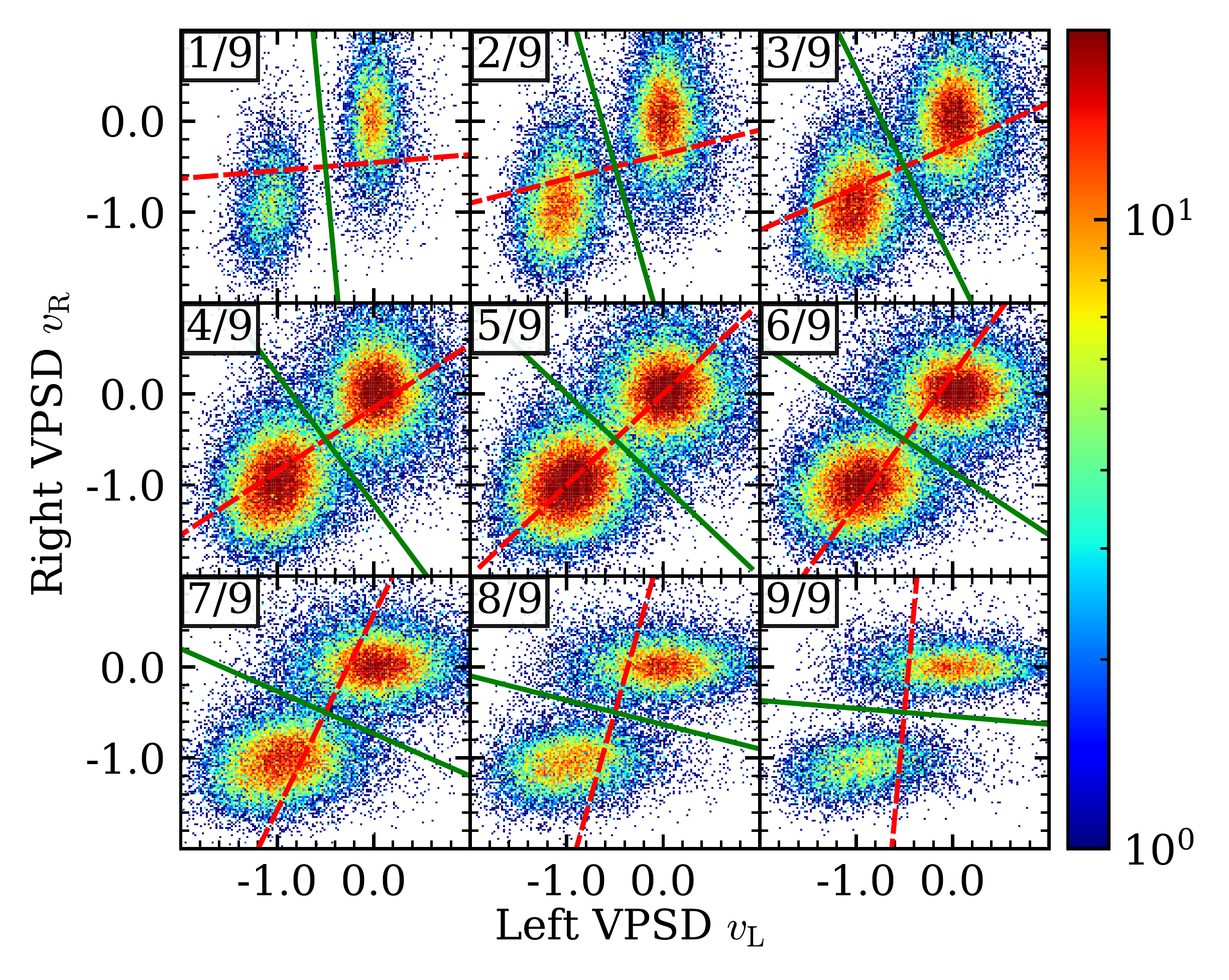}
	\caption{Two-dimensional VPSD plots of the AmBe source radiations for all nine segments of NW bar 01, with $1/9$ represents the leftmost segment and $9/9$ represents the rightmost segment. The solid line represents the n--$\gamma$ separation line, while the dashed line is the line perpendicular to the solid line that pass through $(-0.5,-0.5)$.}
	\label{fig:vpsd_2d_split}
	\end{figure}

	By not using the geometric mean of the signals, one would expect VPSD to exhibit position dependence. In \figurename~\ref{fig:vpsd_2d_split}, the two-dimensional VPSD histogram is split into nine according to their hit positions. Each section spans a length of $\ell/9 = 22\frac{2}{9}~\mathrm{cm}$ of the bar. Counts are not evenly distributed across all segments but instead are maximized around the center ones because the AmBe source was centered along the longitudinal direction of the NW bar. Nonetheless, this non-uniformity of hit position statistics does not hinder a clear pattern from emerging: the closer the hits are to one of the PMTs (see panel $1/9$ and $9/9$), the better the separation of both clusters when projected to the $v_\mathrm{L}$-axis or $v_\mathrm{R}$-axis, respectively.

	By observing how the separation of neutron and gamma clusters evolves as a function of hit position $x\in[-\ell/2,\ell/2]$ in \figurename~\ref{fig:vpsd_2d_split}, we propose a separation line with an angle measured from the horizontal axis as
	\begin{linenomath*}
	\begin{equation}
		\theta_\mathrm{sep}(x) \equiv \frac{\pi}{2}\cdot\left(\frac{x+\ell/2}{\ell}+1\right)  \ ,
	\end{equation}
	\end{linenomath*}
	shown as the solid lines in \figurename~\ref{fig:vpsd_2d_split}. While the two-dimensional VPSD histogram offers more details about the n--$\gamma$ clusters, it is still convenient to quantify ``neutron-ness'' and ``gamma-ness'' with a scalar value. This is done by projecting the clusters onto the dashed lines in \figurename~\ref{fig:vpsd_2d_split} that are defined to be perpendicular to the solid separation lines; both lines intercept at the midpoint between centroids of both clusters, $(-0.5,-0.5)$. Mathematically, we propose a PSD metric named $\mathrm{PPSD}$ (position-corrected VPSD) that is defined as
	\begin{linenomath*}
	\begin{equation}
		\mathrm{PPSD}(x,v_\mathrm{L}, v_\mathrm{R})\equiv\frac{v_\mathrm{L}\cos\theta_\mathrm{proj} + v_\mathrm{R}\sin\theta_\mathrm{proj}}{\cos\theta_\mathrm{proj} + \sin\theta_\mathrm{proj}} \ ,
		\label{eq:PPSD}
	\end{equation}
	\end{linenomath*}
	with the denominator serves to normalize the projected neutron peak to be at $\mathrm{PPSD}\equiv-1$, and
	\begin{linenomath*}
	\begin{equation}
		\theta_\mathrm{proj}(x) \equiv \frac{\pi}{2}\cdot\frac{x+\ell/2}{\ell} \ .
	\end{equation}
	\end{linenomath*}

	\begin{figure}
	\centering
	\includegraphics[width=\columnwidth]{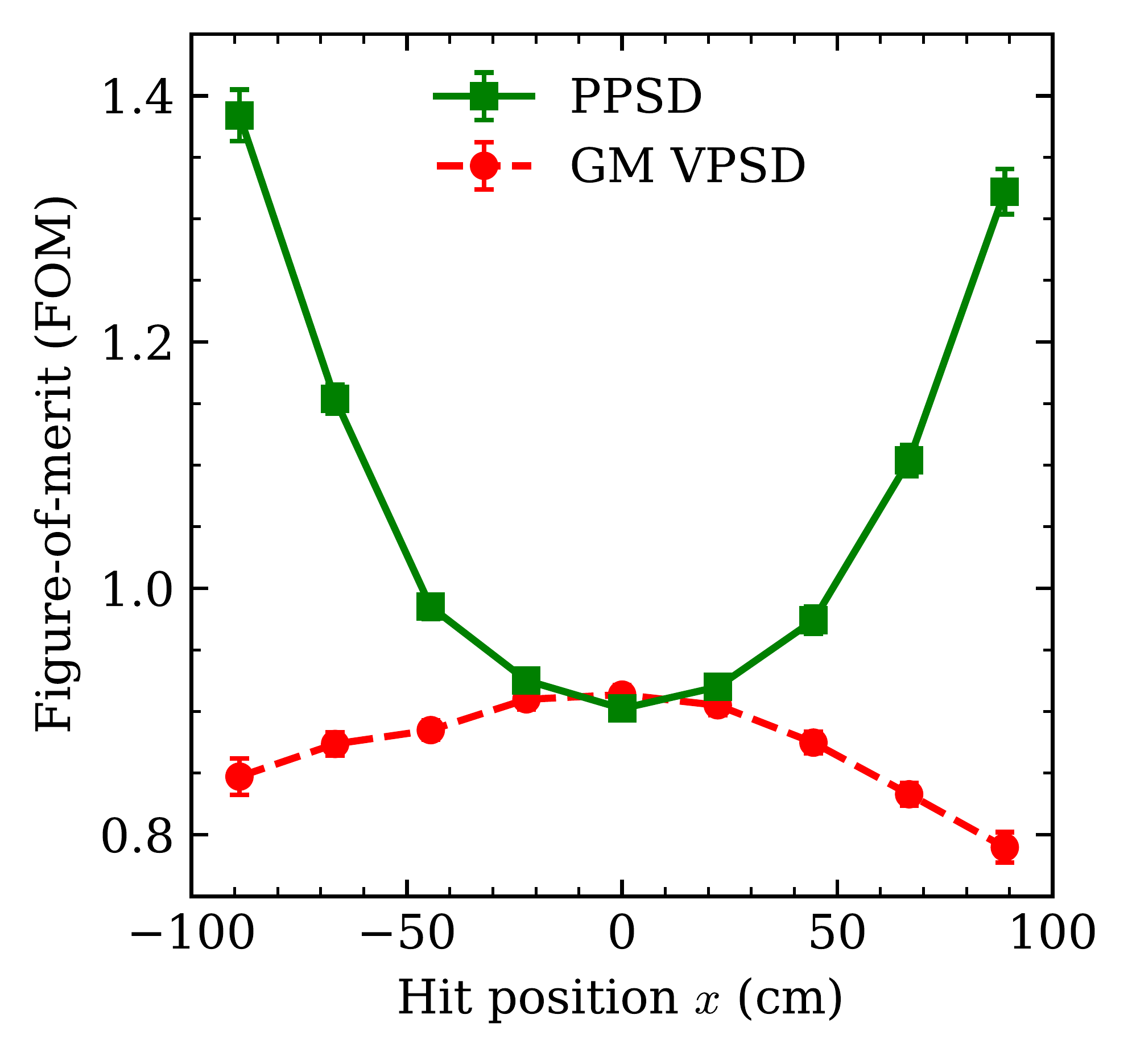}
	\caption{PSD figure-of-merit as a function of hit positions from the AmBe source radiations. Positioned-corrected VPSD (PPSD) significantly improves the n--$\gamma$ discrimination especially close to the edge of a NW bar.}
	\label{fig:figure_of_merit}
	\end{figure}

	Having computed the $\mathrm{PPSD}$ values, we sort them into nine one-dimensional histograms according to their hit positions, and evaluate their respective figure-of-merits (FOMs) by fitting double Gaussian distributions; FOM is defined as
	\begin{linenomath*}
	\begin{equation}
		\mathrm{FOM} = \frac{\left|x_n-x_\gamma\right|}{w_n+w_\gamma}
	\end{equation}
	\end{linenomath*}
	where $x_{n, \gamma}$ is the peak position and $w_{n, \gamma}$ is the corresponding full-width half-maximum (FWHM). To compare the results against the GM method, similar procedure is applied to the VPSD values computed for data points in \figurename~\ref{fig:geomean_2d}, i.e.
	\begin{linenomath*}
	\begin{equation}
		v_\mathrm{GM} \equiv \frac{Q_1^\mathrm{GM}-q_\gamma(Q_2^\mathrm{GM})}{q_\gamma(Q_2^\mathrm{GM}) - q_n(Q_2^\mathrm{GM})} \ .
	\end{equation}
	\end{linenomath*}
	The results are summarized in \figurename~\ref{fig:figure_of_merit}, showing that the overall FOM obtained using the PPSD (solid circles) is superior to those calculated from the GM method (solid squares) especially near the edge of the bar. Signals originating from the far end tend to suffer from a significant reduction in signal-to-noise ratio, and taking the geometric mean of signals from both ends would mean an overall loss in PSD information. Instead, it is better to take the linear combination of one-sided PSD parameters, e.g. $v_\mathrm{L,R}$ (or even the less accurate charge-integration ratio $Q_1^\mathrm{L,R}/Q_2^\mathrm{L,R}$), and assign with some position-dependent weights as done in (\ref{eq:PPSD}).

\subsection{Gate optimization on digitized waveforms}
	\label{ssec:III-E}
	In many analyses, both the LONG and the SHORT gates include the pulse peak \cite{shim2019performance,lotfi2019optimization,flaska2013influence}. In this subsection, we investigate the gate conditions that optimize the n--$\gamma$ discrimination. We vary two gate timings: (1) the common start time for both SHORT and LONG; (2) the stop time of SHORT. The stop time for LONG is set to be the end of the pulse. In other words, we fix $t^\mathrm{start}_1 \equiv t^\mathrm{start}_2\equiv t^\mathrm{start}$, and $t^\mathrm{stop}_2 \equiv 480~\mathrm{ns}$ which is at the end of each digital sampling cycle. For convenience, we also define a reference time $t^\mathrm{ref}$ for each pulse at its $20\%$ constant fraction timing.

	\begin{figure}
	\centering
	\includegraphics[width=\columnwidth]{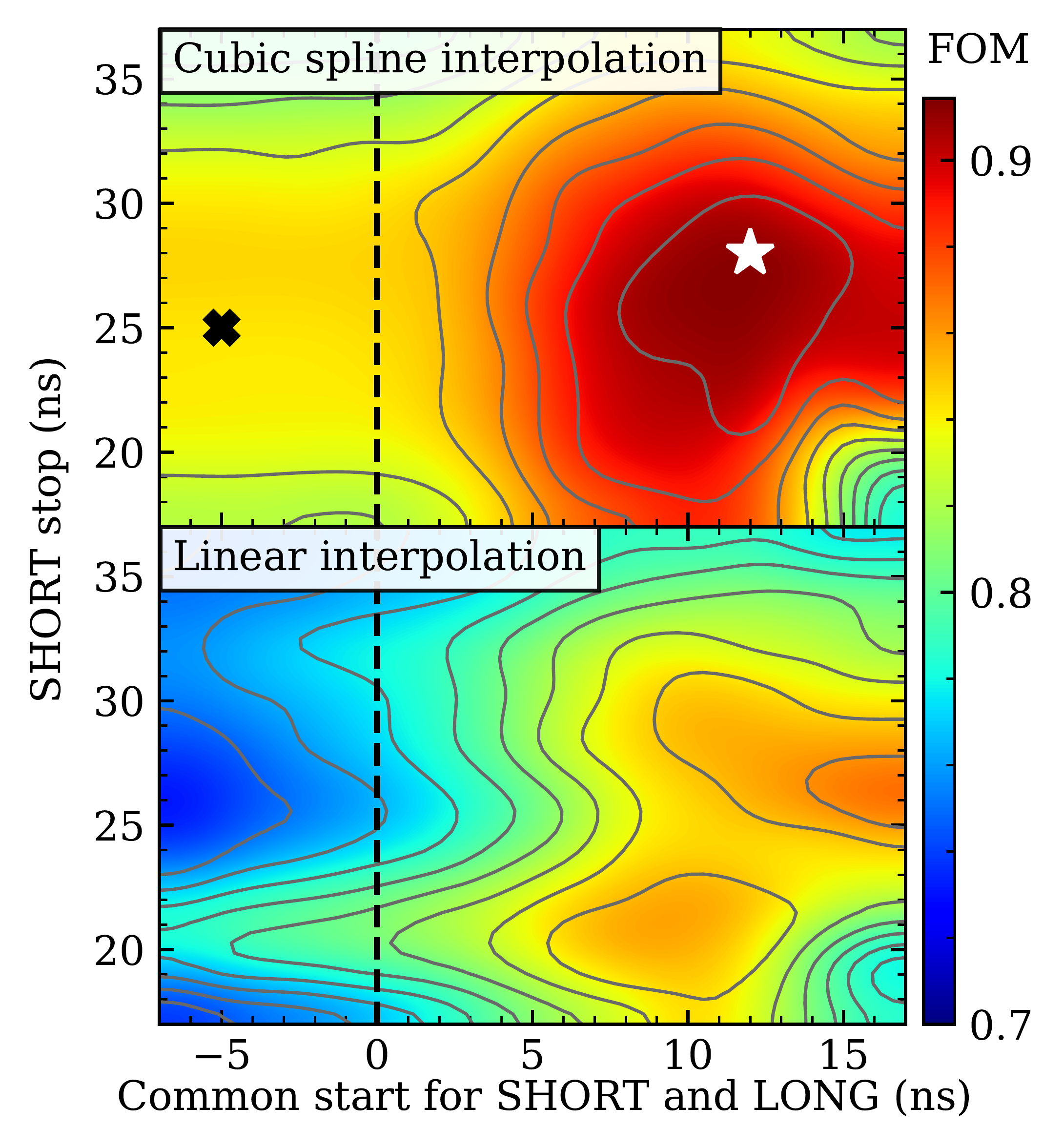}
	\caption{Surfaces of FOM as a function of common gate start and SHORT gate stop with respect to $t^\mathrm{ref}$ (timing at $20\%$) are plotted; curves are contours. Top panel: Cubic spline interpolation on the digitized waveforms yields an optimal FOM of $0.91$ (star marker), whereas the typical gate conditions in an experiment (cross marker) that uses analog electronics measures at a FOM of $0.85$. Bottom panel: Same as top panel but using linear interpolation. Data from the AmBe source.}
	\label{fig:fom_surface}
	\end{figure}

	\begin{figure*}
	\centering
	\includegraphics[width=\textwidth]{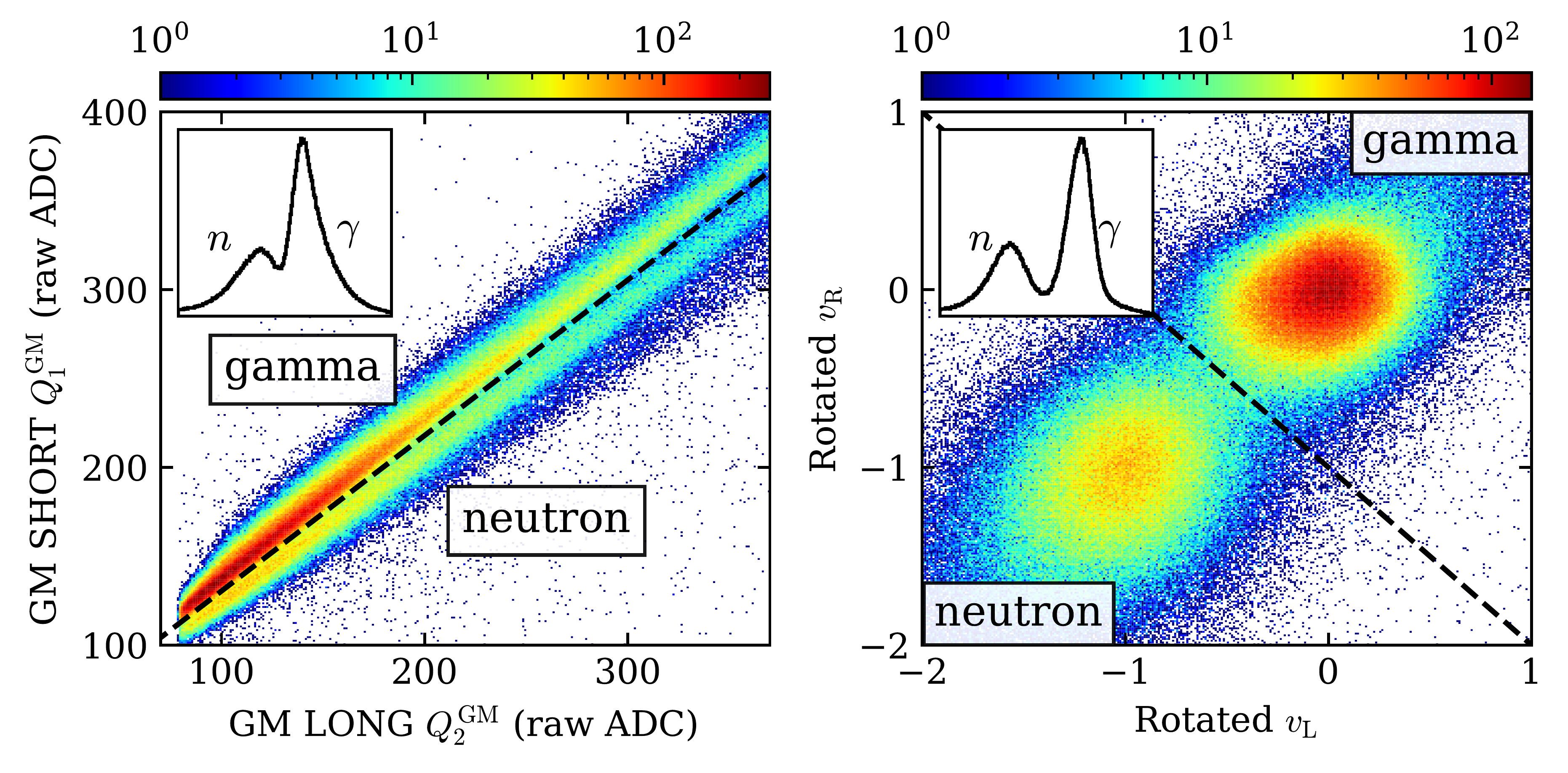}
	\caption{Experimental data on reaction $^{48}\mathrm{Ca}+^{64}\mathrm{Ni}$ at $140~\mathrm{MeV/u}$ using analog processed signals from NW bar 01 using the traditional GM method (left panel) and using the PPSD method (right panel). Both inset figures are projections onto the perpendicular directions of the dashed separation lines.}
	\label{fig:hic_data}
	\end{figure*}

	To select the optimal charge-integration gate conditions, we quantify the n--$\gamma$ separation for every pair of adjustable parameters, namely, the common gate start and the SHORT gate stop. This is done by first projecting all points in the two-dimensional histogram of $(v_\mathrm{L},v_\mathrm{R})$ in \figurename~\ref{fig:vpsd_2d} onto the line connecting both centroids. Then FOM is evaluated by fitting double Gaussian distribution to the one-dimensional projection. The evaluations are done for both cubic spline interpolated and linearly interpolated waveforms, as summarized respectively in the top and bottom panels of \figurename~\ref{fig:fom_surface}. Cubic spline interpolation not only yields a better maximum FOM of $0.91$ (star marker), but also offers an optimal gate condition that is numerically more stable than that of linear interpolation. The optimal FOM using cubic spline interpolation is attained by a common gate start of $t^\mathrm{ref} + 12~\mathrm{ns}$ and a SHORT gate stop of $t^\mathrm{ref} + 28~\mathrm{ns}$. Note that the optimal common gate start is $7~\mathrm{ns}$ after the typical pulse peaks that occur at around $t^\mathrm{ref} + 5~\mathrm{ns}$.

	Since recovery of lost signal information is impossible in analog electronics, caution is exercised in choosing gates. The gates traditionally include the peak partly because light output information is derived from the integrated charge in the LONG pulse. In LANA experiments that use analog electronics, $t^\mathrm{start} = t ^\mathrm{ref} - 5~\mathrm{ns}$ and $t^\mathrm{stop} = t^\mathrm{ref} + 25~\mathrm{ns}$ are chosen. This choice is not optimal but it still yields a FOM of $0.85$ (cross marker) using cubic spline interpolation, being only $6.6\%$ less than the optimal FOM of $0.91$. In other words, the optimal gates which are difficult to achieve in analog electronics are not the most critical factor in PSD performance as long as sensible choice of SHORT and LONG gates is adopted.

\subsection{Experimental data using analog system}
	\label{ssec:hic}

	We have applied the new position-corrected VPSD (PPSD) procedure to the data from a recent experiment with much improvement in n--$\gamma$ discrimination. The analog system was used to process signals from all $25$ NW bars and provide $12$-bit fast gate and total gate information. The fast gate was adjusted to a width of $30~\mathrm{ns}$ that starts from around $5~\mathrm{ns}$ before the $20\%$ threshold timing, while the total gate was set to a width of $340~\mathrm{ns}$. Neither gates were set to the optimal settings as suggested by the star marker in \figurename~\ref{fig:fom_surface}. Nonetheless, as we can see from the FOM surface, the loss in n--$\gamma$ separability is less than $10\%$.

	Data from the heavy-ion collisions of $^{48}\mathrm{Ca} + ^{64}\mathrm{Ni}$ at a beam energy of $140~\mathrm{MeV/u}$ are used to test the new PPSD method. With a veto wall placed in front of the NW bars, charged particles are ``vetoed'' from this analysis, leaving around half a million neutrons and gamma rays detected by NW bar 01. Significant improvement in n--$\gamma$ separation has been presented by comparing the PSD plots produced using the GM method (left panel) and the new PPSD method (right panel) in \figurename~\ref{fig:hic_data}. In order to preserve the two-dimensionality of the clusters in the right panel for better visualization, we rotate each point $(v_\mathrm{L}, v_\mathrm{R})$ by $(\frac{\pi}{4}-\theta_\mathrm{proj}(x))$ around the point $(-0.5, -0.5)$. One-dimensional projections are presented in the respective inset figures. It is clear that the PPSD method exhibits superior n--$\gamma$ separation. Moreover, this new method accommodates many simple clustering algorithms such as K-means clustering \cite{hartigan1979algorithm} that can facilitate automations in the analysis.

\section{Summary}
	Using cubic spline interpolated digitized waveforms from a digital electronic system, we study the properties of a $2~\mathrm{m}$ long NE-213 liquid scintillator neutron detector coupled with a photomultiplier tube (PMT) at each end. In offline analysis, the digitized waveforms allow optimization on the constant fraction of the timing signals, as well as widths and start times of the LONG and SHORT gates for n--$\gamma$ pulse shape discrimination (PSD). The study validates the gating conditions used in analog electronics, that even though they might not be optimal, they do not impair the n--$\gamma$ PSD significantly. The ``valued-assigned PSD'' (VPSD) is used to quantify the separation between neutron and gamma rays in both the left and right PMTs. Position dependence of the light signals is corrected by taking the weighted average of the left and the right VPSD values, with weights linearly dependent on position. Compared to the traditional technique that uses the geometric mean (GM) of light outputs from both PMTs, the PSD figure-of-merit (FOM) increases from $0.8$ to $1.4$ near the edges of the detector. The resulting position-corrected VPSD (PPSD) plot is much improved and allows for easy separation of n and $\gamma$. Finally, we successfully apply PPSD to analyze data from a recent heavy-ion collision experiment.


\vfill

\end{document}